\def\be{\begin{equation}}

\def\ee{\end{equation}}
\def\ba{\begin{array}}
\def\ea{\end{array}}

\def\p{\prime}

\def\Cb{{\Bbb C}}

\documentclass[12pt]{article}
\usepackage{amssymb}
\begin{document}

\setcounter{page}{1}
\centerline{\Large\bf A Note on Equivalence
of Bipartite States}
\bigskip

\centerline{\Large\bf under Local Unitary Transformations}
\vspace{3ex}
\begin{center}

Bao-Zhi Sun$^a$,\, Shao-Ming Fei$^{a,b}$,\, Xianqing Li-Jost$^b$ and Zhi-Xi Wang$^a$
\bigskip
\bigskip

\begin{minipage}{5.2in}
$^a$ Department of Mathematics, Capital Normal University, Beijing 100037\\
$^b$ Max-Planck-Institute for Mathematics in the Sciences, 04103 Leipzig\\

\end{minipage}
\end{center}

\vskip 1 true cm
\parindent=18pt
\parskip=6pt
\begin{center}
\begin{minipage}{5in}
\vspace{3ex} \centerline{\large Abstract} \vspace{4ex}

The equivalence of arbitrary dimensional bipartite states under local
unitary transformations (LUT) is studied. A set of invariants and ancillary
invariants under LUT is presented. We show that two states are equivalent under
LUT if and only if they have the same values
for all of these invariants.

\bigskip
\bigskip

\vfill
\vfill
\smallskip

\vfill

\end{minipage}
\end{center}
\bigskip
\bigskip

The quantum entangled states have been used as the key resources in
quantum information processing and quantum
computation \cite{nielsen}. An important property of
quantum entanglement is that the entanglement of a bipartite
quantum state remains invariant
under local unitary transformations on the subsystems.
Therefore invariants of local unitary transformations
have special importance. For instance
the trace norms of realigned or partial transposed density
matrices in entanglement measure, separability criteria
are some of these invariants \cite{norm}.
Two quantum states are
locally equivalent if and only if all the invariants have equal
values for these two states. For bipartite mixed states,
a generally non-operational method has
been presented to compute all the invariants of local unitary
transformations in \cite{Rains,Grassl}.
In \cite{Makhlin}, the invariants for general two-qubit systems are
studied and a complete set of 18 polynomial invariants is
presented. In \cite{Linden99}, the invariants for three qubits states have been
discussed. A complete set of invariants for generic mixed states
are presented. In \cite{3qubit} the invariants for a class of non-generic
three-qubit states have been investigated.
In \cite{goswami}, complete sets of invariants for
some classes of density matrices have been presented.
The invariants for tripartite pure states have been also studied \cite{wl}.

In \cite{Albeverio03} a complete set of invariants for generic
density matrices with full rank has been presented. In this note we
extend the results to generalized generic density matrices with
arbitrary rank, by taking into account the vector space corresponding to
the zero eigenvalues.

Let $H$ be an $N$-dimensional complex Hilbert space,
with $\vert i\rangle$, $i=1,...,N$, as an orthonormal basis. Let $\rho$ be a
density matrix defined on $H\otimes H$ with $rank(\rho)=n\leq
N^2$. It can be written as
$$
\rho=\sum_{i=1}^n\lambda_i|v_i><v_i|,
$$
where $|v_i>$ is the
eigenvector with respect to the nonzero eigenvalue $\lambda_i$.
$|v_i>$ has the form:
$$
|v_i>=\sum_{k,l=1}^Na_{kl}^i|kl>,\ \ a_{kl}^i\in {\mathcal{C}},\
\ \sum_{k,l=1}^Na_{kl}^ia_{kl}^{i*}=1,\ \ i=1,\cdots,n.
$$
Let $A_i$ denote the matrix given by $(A_i)_{kl}=a_{kl}^i$. We
introduce $\{\rho_i\},\ \{\theta_i\}$,
\begin{equation}
\rho_i=Tr_2|v_i><v_i|=A_iA_i^{\dag},\ \
\theta_i=Tr_1|v_i><v_i|=A_i^tA_i^*,\ i=1,2,\cdots,n,
\end{equation}
where $Tr_1$ and $Tr_2$ stand for the traces over the first and
second Hilbert spaces, $A^t,\ A^*$ are the
transpose and the complex conjugation of $A$ respectively.

Two density matrices $\rho$ and $\rho^\prime$ are said to be equivalent
under local unitary transformations if there exist unitary operators
$U_1$ (resp. $U_2$) on the first (resp. second) space of $H\otimes H$ such that
\be\label{eq}
\rho^\prime=(U_1\otimes U_2)\rho(U_1\otimes U_2)^\dag.
\ee

Let $\Omega(\rho)$ and $\Theta(\rho)$ be two ``metric tensor" matrices,
with entries given by
\begin{equation}\label{omegatheta}
\Omega(\rho)_{ij}=Tr(\rho_i\rho_j),\ \
\Theta(\rho)_{ij}=Tr(\theta_i\theta_j),\ \ \mbox{for}\
i,j=1,\cdots,n.
\end{equation}

We call a mixed state $\rho$ a generic one if $\Omega,\ \Theta$ satisfy
\begin{equation}\label{generic}
\mbox{det}(\Omega(\rho))\neq0\ \ \mbox{and}\ \
\mbox{det}(\Theta(\rho))\neq0.
\end{equation}
In \cite{Albeverio03} it has been shown that two full-ranked bipartite states ($n=N^2$)
satisfying (\ref{generic}) are equivalent under local unitary transformations if and only if they have the
same values of the following invariants:
$\Omega$, $\Theta$,
\begin{equation}\label{xy}
X(\rho)_{ijk}=Tr(\rho_i\rho_j\rho_k),\ \ \
Y(\rho)_{ijk}=Tr(\theta_i\theta_j\theta_k),\ \ i,j,k=1,\cdots,n,
\end{equation}
together with the condition
\begin{equation}\label{js}
J^s(\rho)=Tr(\rho^s),\ s=1,2,\cdots,N^2,
\end{equation}
which guarantee that the density matrices have the same set of eigenvalues.

For the case $n<N^2$, the invariants (\ref{omegatheta}), (\ref{xy}) and (\ref{js})
are no longer enough to verify the equivalence of two generic states
under local unitary transformations, and some ancillary
invariants are needed.

From the generic condition $\mbox{det}(\Omega(\rho))\neq0\ \ \mbox{and}\ \
\mbox{det}(\Theta(\rho))\neq0$, we have that
$\{\rho_i,\,i=1,\cdots,n\}$ and $\{\theta_i,\,i=1,\cdots,n\}$ are two
sets of linear independent matrices.
One can always find some $N\times N$ matrices $\rho_i,\
\theta_i$ (we call them ancillary matrices), $i=n+1,\cdots,N^2$, such that
$\{\rho_i,\,i=1,\cdots,N^2\}$ and $\{\theta_i,\,i=1,\cdots,N^2\}$
span the $N^2\times N^2$ matrix space respectively. Therefore the
$N^2\times N^2$ matrices $\tilde{\Omega}(\rho)$ and $\tilde{\Theta}(\rho)$,
\begin{equation}\label{ot}
\tilde{\Omega}(\rho)_{ij}=Tr(\rho_i\rho_j),\
\tilde{\Theta}(\rho)_{ij}=Tr(\theta_i\theta_j),\
i,j=1,\cdots,N^2,
\end{equation}
satisfy
\begin{equation}\label{ggeneric}
\mbox{det}(\tilde{\Omega}(\rho))\neq0\
\mbox{and}\ \mbox{det}(\tilde{\Theta}(\rho))\neq0.
\end{equation}

Set
\begin{equation}\label{txy}
\tilde{X}(\rho)_{ijk}=Tr(\rho_i\rho_j\rho_k),\ \ \
\tilde{Y}(\rho)_{ijk}=Tr(\theta_i\theta_j\theta_k),\ \
i,j,k=1,\cdots,N^2.
\end{equation}

We call $\tilde{\Omega}(\rho)_{ij}$, $\tilde{\Theta}(\rho)_{ij}$,
$\tilde{X}(\rho)_{ijk}$, $\tilde{Y}(\rho)_{ijk}$, with at least one of
their sub-indices taking values from $n+1$ to $N^2$, the ancillary invariants.

{\sf [Theorem]:} Two generic density matrices are equivalent under local
unitary transformations if and only if there exists a
ordering of the corresponding eigenstates such that the following
invariants have the same values for both density matrices:
\begin{equation}\label{theorem}
J^s(\rho)=Tr(\rho^s),\ s=1,2,\cdots,N^2,\ \
\tilde{\Omega}(\rho),\ \ \tilde{\Theta}(\rho),\ \ \tilde{X}(\rho),\ \ \tilde{Y}(\rho).
\end{equation}

{\sf [Proof]:} Suppose that $\rho$ and $\rho'$ are equivalent under local
unitary transformation $\mu\otimes\omega$,
$\rho'=\mu\otimes\omega\rho\mu^{\dag}\otimes\omega^{\dag}$.
Correspondingly, we have $|\nu_i'>=\mu\otimes\omega|\nu_i>$, i.e.
$A_i'=\mu A_i\omega^t$ for $i=1,...,n$.
As to the ancillary invariants, if $\rho_i,\theta_i,i=n+1,\cdots,N^2$
are the ancillary matrices associated to $\rho$, we can choose
$\rho_i'=\mu\rho_i\mu^{\dag},\
\theta_i'=\omega\theta_i\omega^{\dag},\ i=n+1,\cdots,N^2$ for
$\rho'$.
Therefore
$\rho_i'=A_i'A_i'^{\dag}=\mu\rho_i\mu^{\dag}$,
$\theta_i'=A_i^tA_i^*=\omega\theta_i\omega^{\dag}$, $i=1,\cdots,N^2$.
It is straightforward to verify that the quantities in (\ref{theorem})
are invariants under local unitary transformations, e.g.
$\Omega(\rho')_{ij}=Tr(\rho'_i\rho'_j)=Tr(\mu\rho_i\rho_j\mu^{\dag})
=Tr(\rho_i\rho_j)=\Omega(\rho)_{ij}$,
$\Theta(\rho')_{ij}=Tr(\theta'_i\theta'_j)=Tr(\omega\theta_i\theta_j\omega^{\dag})
=\Theta(\rho)_{ij}$, $i,j=1,\cdots,N^2$.

Conversely we suppose that the states
$\rho=\sum_{i=1}^n\lambda_i\vert \nu_i\rangle\langle\nu_i\vert$ and
$\rho^\prime= \sum_{i=1}^n\lambda_i^\prime\vert\nu_i^\prime\rangle
\langle\nu_i^\prime\vert$ give the same values to the invariants in
(\ref{omegatheta}) and (\ref{xy}). And there exist ancillary
matrices $\rho_i,\rho_i',\theta_i,\theta_i',i=n+1,\cdots,N^2$, such
that they have the same values of (\ref{js}) and the ancillary
invariants in (\ref{ot}) and (\ref{txy}). $\rho$ and $\rho^\prime$
can be proved to be equivalent under local unitary transformations
by using the method in \cite{Albeverio03}. Having the same values of
(\ref{js}) implies that $\rho^\prime$ and $\rho$ have the same
nonzero eigenvalues, $\lambda^\prime_i=\lambda_i$, $i=1,...,n$. The
condition (\ref{ggeneric}), $det(\tilde{\Omega}(\rho))\neq 0$
implies that $\{\rho_i\}$, $i=1,...,N^2$, span the space of $N\times
N$ matrices and therefore \be\label{a} \rho_i\rho_j=\sum_{k=1}^n
C_{ij}^k\rho_k,~~~~~ C_{ij}^k~\in\Cb, \ee which gives rise to
$\tilde{\Omega}_{ij}=\sum_{k=1}^n C_{ij}^k$. Hence
$\tilde{X}_{ijk}=\sum_{l=1}^n C_{ij}^l\tilde{\Omega}_{lk}$ and
\be\label{d}
C_{ij}^l=\sum_{k=1}^n\tilde{X}_{ijk}\tilde{\Omega}^{lk}, \ee where
the matrices $\tilde{\Omega}^{ij}$ is the corresponding inverses of
the matrices $\tilde{\Omega}_{ij}$. We have that $\{\rho_i\}$ form
an irreducible $N$-dimensional representation of the algebra
$gl(N,\Cb)$ with structure constants $C_{ij}^k-C_{ji}^k$. Similarly
$\{\rho^{\prime}_i\}$, $i=1,...,N^2$, also form an irreducible
$N$-dimensional representation of the algebra $gl(N,\Cb)$ with same
structure constants. These two sets of representations of the
algebra $gl(N,\Cb)$ are equivalent, $\rho_i^\p=u\rho_i u^\dag$, for
some unitary $u$.

Similarly, from $\tilde{\Theta}(\rho)=\tilde{\Theta}(\rho^{\prime})$
and $\tilde{Y}_{ijk}(\rho)=\tilde{Y}_{ijk}(\rho^{\prime})$ we can
deduce that $\theta_i^\p=w^{\dag}\theta_i w$, for some unitary $w$.
From the singular value decomposition of matrices, we have
$\vert\nu_i^\prime\rangle=u\otimes w \vert\nu_i\rangle$,
$i=1,...,n$, and $\rho^\prime=u\otimes w ~\rho~u^\dag\otimes
w^\dag$. Hence $\rho^\prime$ and $\rho$ are equivalent under local
unitary transformations. \hfill $\rule{2mm}{2mm}$

{\sf Remark} The invariants in (\ref{theorem}) could be redundant.
For example when $\rho$ and $\rho'$ are $2\times 2$ pure states, one
only needs $Tr\rho_1\rho_1=Tr\rho'_1\rho'_1$ for verifying the
equivalence of them. But for higher dimensional systems, all these
invariants are needed for generic states. Moreover when $\rho$ and
$\rho'$ has equal nonzero eigenvalues, if they are equivalent under
local unitary transformations we can always find a set of
eigenvectors suitably labeled such that they have the same
invariants (\ref{theorem}), as seen from the proof.

As an example, let
$$|\psi_1>=\frac{1}{\sqrt{2}}(|00>+|11>),\
|\psi'_1>=\frac{1}{\sqrt{2}}(|01>+|10>)
$$
and
$$|\psi_2>=|01>,\ \ \ \ \ |\psi_2'>=|00>.$$
We consider
$$\rho=\frac{1}{3}|\psi_1><\psi_1|+\frac{2}{3}|\psi_2><\psi_2|,$$
$$\rho'=\frac{1}{3}|\psi'_1><\psi'_1|+\frac{2}{3}|\psi_2'><\psi_2'|.$$
These matrices have the same eigenvalues.
The corresponding eigenvectors give rise to
$$\rho_1=\theta_1=\rho_1'=\theta_1'=\left(\matrix{\frac{1}{2}&0\cr
0&\frac{1}{2}}\right),$$
$$\rho_2=\rho'_2=\theta'_2=\left(\matrix{1&0\cr0&0}\right),\
\ \theta_2=\left(\matrix{0&0\cr0&1}\right).$$
By directly calculations we have,
$$\Omega(\rho)=\Omega(\rho')=\left(\ba{cc}\frac{1}{2}&\frac{1}{2}\\[2mm]
\frac{1}{2}&1\ea\right)=\Theta(\rho)=\Theta(\rho')$$
$$\begin{array}{l}X(\rho)_{ijk}=X(\rho')_{ijk}
=Y(\rho)_{ijk}=Y(\rho')_{ijk}\\
\\
\ \ \ \ \ \ \ \ =\left\{
\begin{array}{l}1,\ \mbox{if}\ \ i=j=k=2\\[2mm]
\frac{1}{4},\ \mbox{if}\ \ ijk\in\{111,112,121,211\}\\[2mm]
\frac{1}{2},\ \mbox{for the rest}.\end{array}\right.\end{array}$$
So $\mbox{det}(\Omega)=\mbox{det}(\Theta)=\frac{1}{2}\neq0$
for both $\rho,\ \rho'$, and they are generic states.
We choose the ancillary matrices as:
  $$\rho_3=\rho_3'=\left(\matrix{0&1\cr
 0&0}\right)=\theta_3=\theta'_4,\ \rho_4=\rho_4'=\left(\matrix{0&0\cr
 1&0}\right)=\theta_4=\theta'_3.$$
We have
 $$\tilde{\Omega}(\rho)=\tilde{\Omega}(\rho')
 =\left(\matrix{
 \frac{1}{2}&\frac{1}{2}&0&0\cr \frac{1}{2}&1&0&0\cr
 0&0&0&1\cr 0&0&1&0}\right)
 =\tilde{\Theta}(\rho)=\tilde{\Theta}(\rho')$$
 and
 $$\tilde{X}(\rho)_{ijk}=\tilde{X}(\rho')_{ijk}=
 \left\{\begin{array}{l l}X(\rho)_{ijk},\ \ &i,j,k=1,2\\[2mm]
 \frac{1}{2},\ \ &ijk\in\{134,341,413,143,431,314\}\\[2mm]
 1,\ \ &ijk\in\{234,342,423\}\\[2mm]
 0,\ \ &\mbox{for else}\end{array}\right.,$$
 $$\tilde{Y}(\rho)_{ijk}=\tilde{Y}(\rho')_{ijk}=
 \left\{\begin{array}{l l}X(\rho)_{ijk},\ \ &i,j,k=1,2\\[2mm]
 \frac{1}{2},\ \ &ijk\in\{134,341,413,143,431,314\}\\[2mm]
 1,\ \ &ijk\in\{243,432,324\}\\[2mm]
 0,\ \ &\mbox{for else}\end{array}\right..$$
From the theorem we can conclude that $\rho$ and $\rho'$ are
equivalent under local unitary transformations.

We have studied the equivalence of two bipartite states with
arbitrary dimensions by using some ancillary invariants under the
unitary transformation. This method applies to all the
bipartite generic density matrices.
As for the nongeneric states, i.e. the states
satisfying $\mbox{det}(\Omega(\rho))=0$ or $\mbox{det}(\Theta(\rho))=0$
or both, we can deal with the problem in the following way.
If $\mbox{det}(\Omega(\rho))=0$, i.e. $\rho_i,\ i=1,\cdots,n$, are linear dependent,
we choose the maximal linear
independent subset (denote as $S$) of $\{\rho_i,\ i=1,\cdots,n\}$.
For two states $\rho$ and $\rho'$ with the same invariants in
(\ref{theorem}), we only need to find a unitary matrix $\mu$ satisfying
$\rho_i'=\mu\rho_i\mu^{\dag}$ for $\rho_i\in S$ and $\rho'_i\in
S'$. According to the subset $S$, we can get submatrices of
$\Omega(\rho)$ and $X(\rho)$. We denote these submatrices as
$\bar{\Omega}(\rho),\ \bar{X}(\rho)$. We extend them to matrices
$\tilde{\Omega}(\rho),\ \tilde{\Theta}(\rho)$ instead of
$\Omega(\rho),\ X(\rho)$. Then using the theorem and the relation
between $S$ (resp. $S'$) and $\{\rho_i,i=1,\cdots,n\}$ (resp.
$\{\rho'_i,i=1,\cdots,n\}$), we can get
$\rho_i'=\mu\rho_i\mu^{\dag}, i=1,\cdots,n$ for some unitary
matrix $\mu$. The case of
$\mbox{det}(\Theta(\rho))=0$ can be similarly treated.

\end{document}